\documentclass[]{aa} 
\usepackage{graphicx,rotating}
\usepackage{txfonts}
\usepackage{soul}
\usepackage{natbib}
\usepackage{subfig}
\usepackage{color}
\begin{document}
\title{Stratified NH and ND emission in the prestellar core 16293E in L1689N}
\titlerunning{NH and ND in the prestellar core 16293E}
\authorrunning{Bacmann et al.}
\author{
A. Bacmann\inst{\ref{inst0},\ref{inst1}} 
\and F.~Daniel\inst{\ref{inst0},\ref{inst1}} 
\and P. Caselli\inst{\ref{inst2}}
\and C. Ceccarelli\inst{\ref{inst0},\ref{inst1}} 
\and D. Lis\inst{\ref{inst3},\ref{inst4}}
\and C. Vastel\inst{\ref{inst5},\ref{inst6}}
\and F. Dumouchel\inst{\ref{inst7}} 
\and F. Lique\inst{\ref{inst7}} 
\and E. Caux\inst{\ref{inst5},\ref{inst6}}
}
\institute{
Univ. Grenoble Alpes, IPAG, F-38000 Grenoble, France  \label{inst0} 
\and
CNRS, IPAG, F-38000 Grenoble, France \label{inst1} 
\and
 Max Planck Institute for Extraterrestrial Physics, Giessenbachstrasse 1, D-85748 Garching, Germany \label{inst2} 
\and
LERMA, Observatoire de Paris, PSL Research University, CNRS, Sorbonne Universit\'es, UPMC Univ. Paris 06, 75014 Paris, France\label{inst3}
\and
California Institute of Technology, Cahill Center for Astronomy and Astrophysics 301-17, Pasadena, CA 91125, USA  \label{inst4} 
\and
Universit\'e de Toulouse; UPS-OMP; Institut de Recherche en Astrophysique et Plan\'etologie (IRAP), Toulouse, France \label{inst5}
\and
CNRS, IRAP, 9 Av. colonel Roche, BP 44346, 31028 Toulouse Cedex 4, France \label{inst6}
\and 
LOMC-UMR 6294, CNRS-Universit\'e du Havre, 25 rue Philippe Lebon, BP 1123 -- 76063 Le Havre Cedex, France \label{inst7} 
}

\date{Received; accepted}


\abstract
{High degrees of deuterium fractionation are commonly found in cold pre-stellar cores and in the envelopes around young protostars. As it brings strong constraints to chemical models, deuterium chemistry is often used to infer core history or molecule formation pathways. Whereas a large number of observations is available regarding interstellar deuterated stable molecules, relatively little is known about the deuteration of hydride radicals, as their fundamental rotational transitions are at high frequencies where the atmosphere is mostly opaque.}
{Nitrogen hydride radicals are important species in nitrogen chemistry, as they are thought 
to be related to ammonia formation. Observations have shown that ammonia is strongly deuterated, with [NH$_2$D]/[NH$_3]\sim 10$\%. Models predict similarly  high [ND]/[NH] ratios, but so far only one observational determination of this ratio is available, towards the envelope of the protostar IRAS16293-2422. In order to test model predictions, we aim here at determining [ND]/[NH] in a dense, starless core.}
{We observed NH and ND in 16293E with the HIFI spectrometer on board the \textit{Herschel} Space Observatory as part of the CHESS guaranteed time key programme, and derived the abundances of these two species using a non-LTE non-local radiative transfer model.}
{Both NH and ND are detected in the source, with ND in emission and NH in absorption against the continuum arising from the cold dust emission. Our model shows however that the ND emission and the NH absorption originate from different layers in the cloud, as further evidenced by their different velocities. In the central region of the core, we can set a lower limit to the [ND]/[NH] ratio of $\gtrsim$2\%. This estimate is consistent with recent pure gas-phase models of nitrogen  chemistry.}
{}
\keywords{Astrochemistry --- Radiative transfer --- ISM: molecules ---ISM: abundances}

\maketitle
%

\section{Introduction}

Cold and dense regions of the interstellar medium like protostellar envelopes or prestellar cores have long been known to display high degrees of molecular deuterium fractionation \citep[e.g.][]{Jefferts:1973dw}. Despite a low elemental interstellar D/H ratio 
of $\sim 1.5\;10^{-5}$ \citep{Linsky:2006kq}, the abundance ratios of deuterated molecules over their hydrogenated counterparts can reach from a few percent up to several tens of percent, depending on the species and  the source. Doubly and triply deuterated species like CD$_3$OH or ND$_3$ have been detected in cold and dense gas, with fractionation ratios as high as 0.1\% for [ND$_3$]/[NH$_3$] \citep{Lis:2002ft} and 1\% for [CD$_3$OH]/[CH$_3$OH]  \citep{Parise:2004go} the latter representing an enrichment of over 11 orders of magnitude over the [HD]/[H$_2$] ratio. This abundance enhancement in deuterated molecules has its origins in the difference of zero-point energies between H$_3^+$ and its deuterated isotopologue H$_2$D$^+$, so that the exchange reaction:
\begin{equation}
\mathrm{H_3^+ + HD \rightarrow H_2D^+ + H_2}
\label{reaction1}
\end{equation}
is slightly exothermic, and favours the formation of H$_2$D$^+$ \citep{Watson:1976is}. 
The exothermicity of this reaction will depend on the spin--state of the molecular species \citep{Pagani:1992vg,Flower:2006ea,Pagani:2011bl}. As H$_2$ is mainly in its \textit{para} form in dark clouds \citep{Pagani:2009ct}, the reverse reaction does not proceed at low temperature and the formation of H$_2$D$^+$ is favoured. 
Finally, it is only when abundant species that are  H$_3^+$ destroyers, like CO, start freezing out onto the grains that reaction (\ref{reaction1}) becomes the major destruction route for H$_3^+$, increasing the abundance ratio [H$_2$D$^+$]/[H$_3^+$]. H$_2$D$^+$ can then transfer its deuteron to other species via ion-molecule reactions such as: CO + H$_2$D$^+$ $\rightarrow$ DCO$^+$ + H$_2$ \citep{1984ApJ...287L..47D}. Reactions similar to (\ref{reaction1}) involving multiply deuterated isotopologues of H$_3^+$ also contribute to the increase in the molecular D/H ratio \citep{Phillips:2003vu,Roberts:2003en,Walmsley:2004hc}. Analogous mechanisms are thought to be at work with the reactions
\begin{displaymath}
\mathrm{CH_3^+ + HD \leftrightarrows CH_2D^+ + H_2}
\end{displaymath}
\begin{displaymath}
\mathrm{C_2H_2^+ + HD \leftrightarrows C_2HD^+ + H_2}
\end{displaymath}
so that CH$_2$D$^+$ and C$_2$HD$^+$ may contribute to deuteration \citep{Parise:2009co}. These reactions are more exothermic than reaction (\ref{reaction1}) and are therefore believed to be dominant in  warmer environments where   reaction (\ref{reaction1}) is no longer efficient \citep[see][for a comprehensive discussion on CH$_2$D$^+$]{Roueff:2013cd}.

As it depends sensitively on molecular depletion, deuteration can be used to probe the history of molecular cloud formation \citep{Caselli2012,Pagani:2013in,Ceccarelli2014, Brunken:2014ee}. 
Indeed, the molecular D/H is expected to increase with time, as the core condenses and its density increases, and CO depletion sets in. 

Deuterium fractionation in ammonia has been observed to be very high \citep{Roueff:2000tq}. These fractionation ratios were well accounted for by pure gas-phase models \citep{Roueff:2005jh}.   The model of \citet{Roueff:2005jh} also predicts high D/H ratio in nitrogen hydride radicals (e.g. NH, NH$_2$), which are important by-products of ammonia formation. These species are believed to be formed at late stages,  
because they originate from molecular nitrogen N$_2$, which in turn is the product of  (much slower) neutral-neutral chemistry \citep{LeGal:2014dv}.  
  
Observationally, nitrogen hydride radicals have long been elusive in dark cloud cores: because of  their low moments of inertia, these species have their fundamental rotational transitions in the submillimeter to far-infrared range, where the high atmospheric opacity makes ground-based observations at best extremely challenging. The {\it Herschel} Space Observatory with the heterodyne spectrometer HIFI has enabled the observations of hydrides in the 500\,GHz--2\,THz range at an unprecedented sensitivity and spectral resolution. Nitrogen hydride radicals NH and NH$_2$ have been detected in the envelope of the Class 0 protostar IRAS\,16293-2422, where they are seen in absorption against the continuum produced by the warm dust close to the central protostellar source \citep{HilyBlant:2010jp}. Deuterated imidogen, ND, was also detected by \citet{Bacmann:2010it} in the same source, and the derived [ND]/[NH] ratio was found very high (around 70\%).

Deuterated molecules are also used to probe chemical pathways. In the case of nitrogen hydrides, NH is believed to form mostly from the dissociative recombination of N$_2$H$^+$ \citep{Roueff:2015cc, Dislaire:2012bu,HilyBlant:2010jp}, whereas NH$_2$ is a product of the dissociative recombination of hydride ions like NH$_3^+$ and NH$_4^+$, themselves coming from successive hydrogenations of N$^+$ by H$_2$. Deuterium fractionation should reflect the different formation paths, as it is expected to be different depending on the origin of the deuteration. Indeed \citet{Roueff:2005jh} find the [ND]/[NH] ratio larger by a factor of 10 than the [NHD]/[NH$_2$] ratio. 

Many studies of deuterium fractionation are based on the ratio integrated along the line of sight, and make therefore the implicit assumption that the considered molecule and its deuterated counterpart are coexistent. However, conditions required to form the hydrogenated species may be different from those necessary to produce high deuterium fractionations, so that the deuterated species may not coexist everywhere with its hydrogenated counterpart
\citep[see e.g.][]{Caselli:2002cb}.
Here we present {\it Herschel}/HIFI observations of NH and ND in the 16293E prestellar core, a source which is in the vicinity of the Class 0 protostar IRAS16293-2422. The source is known to harbour high abundances of deuterated molecules \citep{Loinard:2001im,Lis:2002jz,Vastel:2004ja}, and to have moderately cold temperatures for a starless core  \citep[around 15\,K $-$][]{Stark:2004hz}, making it easier to detect species in absorption against the dust continuum emission. The observations are described in Section\,\ref{sec:obs} and a first order LTE analysis in Section\,\ref{sec:results}. In Section\,\ref{sec:model} we present our non-LTE non-local radiative transfer modeling of the source and discuss the findings in Section\,\ref{sec:discussion}. In this paper, square brackets refer to abundances with respect to H$_2$.

\section{Observations}\label{sec:obs}

The source 16293E was observed with HIFI  \citep{deGraauw:2010gy} on board the {\it Herschel} Space Observatory \citep{Pilbratt:2010en} as part of the guaranteed time key programme CHESS \citep{Ceccarelli:2010hu}. The ND ($N_ J=0_1-1_2$) transition at 522\,GHz was observed on August 26, 2011 (observation ID 1342227402) and the NH ($N_ J=0_1-1_2$) transition at 974\,GHz on September 15, 2011 (observation ID 1342228623), with both the acousto-optic wide band spectrometer (WBS, spectral resolution 1.1\,MHz, corresponding to a velocity resolution of 0.6\,km/s at 522\,GHz and 0.35\,km/s at 974\,GHz) and the high resolution spectrometer (HRS, spectral resolution 0.25\,MHz, corresponding to 0.15\,km/s at 522\,GHz and 0.08\,km/s at 974\,GHz). The  observing mode used was the dual beam switch with optimisation of the continuum. In this mode  an internal mirror is moved to provide two symmetric OFF positions situated 3$^\prime$ on either side (east and west) of the ON position. For 16293E, one of the OFF positions is close to an outflow driven by the protostar IRAS16293-2422, where we do not expect ND emission. We have checked the OFF positions and found they showed no detectable signal. 

Integration times (ON+OFF) were 80 minutes on ND and nearly 8 hours on NH. The source coordinates for the integration were $\alpha_{2000}=16^h$32$^m$28.6$^s$, $\delta_{2000}=-24{\degr}29^\prime03^{\prime\prime}$. This position corresponds to the peak of the DCO$^+$ emission mapped by \citet{Lis:2002jz}.   The {\it Herschel} full width at half maximum beam sizes were 41$^{\prime\prime}$ at 522\,GHz and 22$^{\prime\prime}$ at 974\,GHz \citep{Roelfsema:2012gb}.

The heterodyne data were processed with the standard HIFI pipeline version 8 up to level 2 products, after which they were exported to Gildas/CLASS\footnote{http://www.iram.fr/IRAMFR/GILDAS} data format. Further analysis consisted of averaging individual spectra in both horizontal and vertical polarisations for the two backends and fitting a straight line to line-free regions in order to remove a baseline. The line brightness was converted to the main beam temperature using the latest 
estimate\footnote{http://herschel.esac.esa.int/twiki/pub/Public/HifiCalibrationWeb/\-HifiBeamReleaseNote\_Sep2014.pdf}
of the forward and beam efficiencies, i.e. $B_{\rm eff}=0.64$ at 522\,GHz, $B_{\rm eff}=0.65$ at 974\,GHz, and $F_{\rm eff}=0.96$.
The rest of the heterodyne data analysis was performed in CLASS.
 
In the frequency ranges where hyperfine components of NH and ND are expected (and corresponding image bands, as our observations are double side-band), we have verified the absence of contamination from lines of other species. Each detected transition in the observed spectrum correspond to a component of the NH or ND transition. Moreover, spectroscopic catalogues (JPL and CDMS) contain very few transitions which would be detectable in the physical conditions prevailing in prestellar cores (i.e. typically with A$_{ij} > 10^{-7}$\,s$^{-1}$ and E$_{\rm up} < 50$\,K for emission lines or E$_{\rm lo}<50$\,K for absorption lines, and these frequencies do not correspond to any detected feature of the spectrum). The rest frequencies and relative intensities of the hyperfine components for the observed NH and ND transitions are listed in Appendix\,\ref{sec:frequencies}.

In addition to the heterodyne data, we also used continuum data in order to take into account molecular excitation by the photons emitted by the dust. Maps of 16293E at 160\,$\mu$m taken from the PACS instrument \citep{Poglitsch:2010bm} aboard {\it Herschel} and at 250, 350, and 500\,$\mu$m from the SPIRE instrument \citep{Griffin:2010hz} were taken as part of the Herschel Gould Belt survey guaranteed time key program \citep{Andre:2010ka}. The PACS data (observation ID 1342241499 and 13422414500) consisted of two maps scanned in orthogonal directions in order to suppress the $1/f$ noise of the detectors.  The data were processed within the HIPE (version 12.0) environment with the Scanamorphos reduction and map reconstruction algorithm \citep{Roussel:2013hp}. In addition, we subtracted from the whole image a constant offset determined from an apparently emission-free area to the South-East of the L1689 cloud. This works as a first order approximation to remove the emission of the telescope. For the SPIRE observations (observation ID 1342205093 and 1342205094), we used the level 2.5 maps  automatically processed by the pipeline and calibrated for extended sources. Absolute calibration using HFI data from the {\it Planck} Satellite are also performed by the pipeline. The angular resolutions of the continuum maps are 18, 25, and 36$\arcsec$ for the SPIRE maps at 250 350, and 500\,$\mu$m \citep[for more details, see][]{Kirk:2013cr}, respectively, and around 14$\arcsec$ for the PACS map. Ground based observations at 850\,$\mu$m from the James Clerk Maxwell 15\,m Telescope  (JCMT) on Mauna Kea were taken with the SCUBA-2 bolometer array \citep{Holland:2013eq}  as part of the JCMT Gould Belt Survey. The data are presented in \citet{Pattle:2015bq} where the data reduction procedure is also described. The 1.3\,mm continuum map was taken at the IRAM 30\,m telescope on Pico Veleta, Spain with the MPIfR bolometer array and is published in \citet{Lis:2002jz}. The beams were 14.1$\arcsec$ for the JCMT map and 11$\arcsec$ for the IRAM 30\,m map.

The distance to the source is taken to be 120 pc \citep{Knude:1998tk,Loinard:2008fy}.

\section{Results}\label{sec:results}

\begin{figure}
\begin{center}
\includegraphics[angle=0,scale=.4]{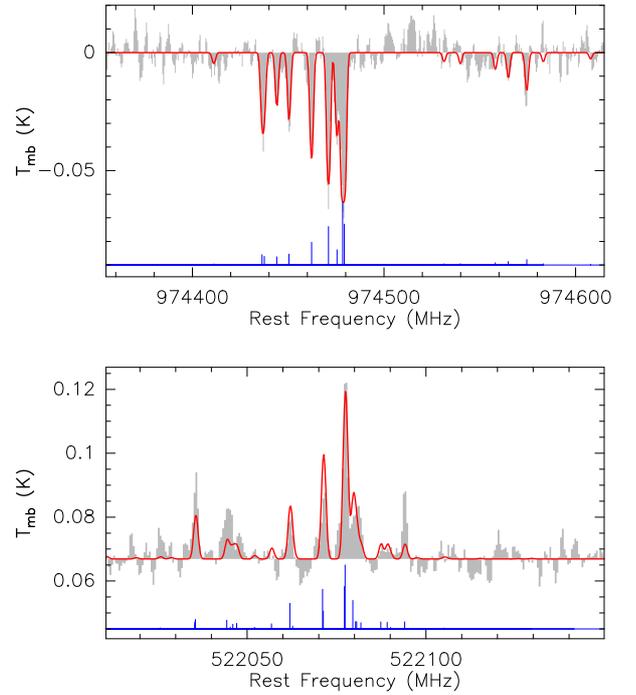}
\caption{Observed spectra of NH (top) and ND (bottom) in grey histograms, with the LTE model as a red solid line. As the continuum is unreliable for NH, it has been subtracted from the spectrum. The LTE fit poorly reproduces the relative line ratios in ND. A schematic view of the hyperfine structure indicating the frequencies and relative intensities of the hyperfine components is shown at the bottom of each spectrum (blue). } \vspace{-0.1cm}
\label{fig:fit_lte}
\end{center}
\end{figure}

The ND transition (lower panel in Fig. \ref{fig:fit_lte}) is detected in emission above a continuum of about 0.067 K (single side band value) and the hyperfine components are mostly resolved. The $rms$ on the spectrum in 1\,MHz channels is less than 3\,mK. The NH transition at 974.5\,GHz (upper panel in Fig. \ref{fig:fit_lte})  is detected in absorption, and its hyperfine components are also mostly resolved. The continuum level however oscillates by several tens of mK over each 1\,GHz spectral window of the WBS and can therefore not be relied on. However, fitting a greybody curve to the continuum taken in the same source at other frequencies with HIFI yields estimates of the continuum around 0.07\,K (see Appendix\,\ref{appendix:SED}).

In a first order analysis, we fitted the hyperfine components of the ND and NH transitions using the HFS method in CLASS, the same way as described in \citet{Bacmann:2010it}. The HFS fit yields the source velocity, the line width, the excitation temperature as well as the line opacities. We fitted both the WBS spectra which have better signal-to-noise but lack spectral resolution as well as the HRS spectra in order to have a better estimate of the line width. The source velocities given by the fits are 3.53$\pm$0.02 km/s for ND and 4.0$\pm0.01$\,km/s for NH, indicating that the NH producing the absorption and the ND producing the emission may not be spatially coexistent. This is reminiscent of the velocity differences seen in this source, where the velocity of the emission from deuterium bearing species like DCO$^+$, N$_2$D$^+$ \citep{Lis:2002jz}, H$_2$D$^+$ and D$_2$H$^+$\citep{Vastel:2004ja,Vastel:2012fd} or ND$_2$H \citep{Gerin:2006ey}, as well as from dense gas tracers like N$_2$H$^+$ \citep{Castets:2001dw} are close to $3.5-3.7$\,km/s whereas those of lower-density tracers \citep[e.g. H$^{13}$CO$^+$, C$^{18}$O and C$^{17}$O - ][]{Lis:2002jz, Stark:2004hz} are at the cloud velocity of $3.8-4$\,km/s. The line widths found by fitting the HRS spectra are 0.4$\pm$0.02\,km/s for  both ND and NH, consistent with line widths measured in this source in other hydrides like ND$_2$H \citep{Lis:2006gw}. 
Finally, we refer the reader to the discussion  in \citet{Vastel:2012fd} where more details on these various points can be found.
For ND, the hyperfine components ratios are poorly reproduced in the LTE approach, as illustrated in Fig.\,\ref{fig:fit_lte}.  This indicates non-LTE effects (including, for example, excitation effects due to component overlap) which can arise in the case of high opacities. The opacities found by the fit in the case of ND are likely underestimated, as they are found quite low ($\sim 1.7$ for the total opacity summed over all hyperfine components). The column densities for ND and NH using the LTE approximation are 1.1 10$^{13}\pm 0.9\;10^{13}$cm$^{-2}$ and 5.8 10$^{13}\pm 1.3\;10^{13}$ cm$^{-2}$, respectively. The method used to derive these values is described in more details in \citet{Bacmann:2010it}. The NH column density is found not to be sensitive to the value of the continuum intensity for continuum temperatures below 0.5\,K. This is due to the large energy difference between the J=0 and J=1 levels so that the J=1 level  is not populated and $h\nu\gg k T_{ex}$.
The excitation temperature for NH (as derived from the HFS fit) is found to be around 5\,K if we assume the continuum to be 0.07\,K, and the total opacity (summed over all hyperfine components) is 12.5. 

\section{Radiative transfer modelling} \label{sec:model}

Because of the poor fit of the hyperfine component ratios in ND and the velocity difference between ND and NH, we decided to take advantage of the new NH and ND collisional data recently published \citep{Dumouchel:2012fl} and used a more comprehensive non-LTE radiative transfer model in order to reproduce both spectral and continuum observations. The radiative transfer model takes into account line overlap for excitation  as well as possible radiative excitation by dust. A detailed description of the code can be found in \citet{Daniel:2008fy}.

\subsection{Continuum}\label{sec:continuum}

We first built a model of the physical structure
of the 16293E core, based on continuum observations at 160, 250, 350, 500, 850 $\mu$m and 1.3\,mm (see Sect. \ref{sec:obs}).

\begin{figure}
\begin{center}
\includegraphics[angle=0,scale=.45]{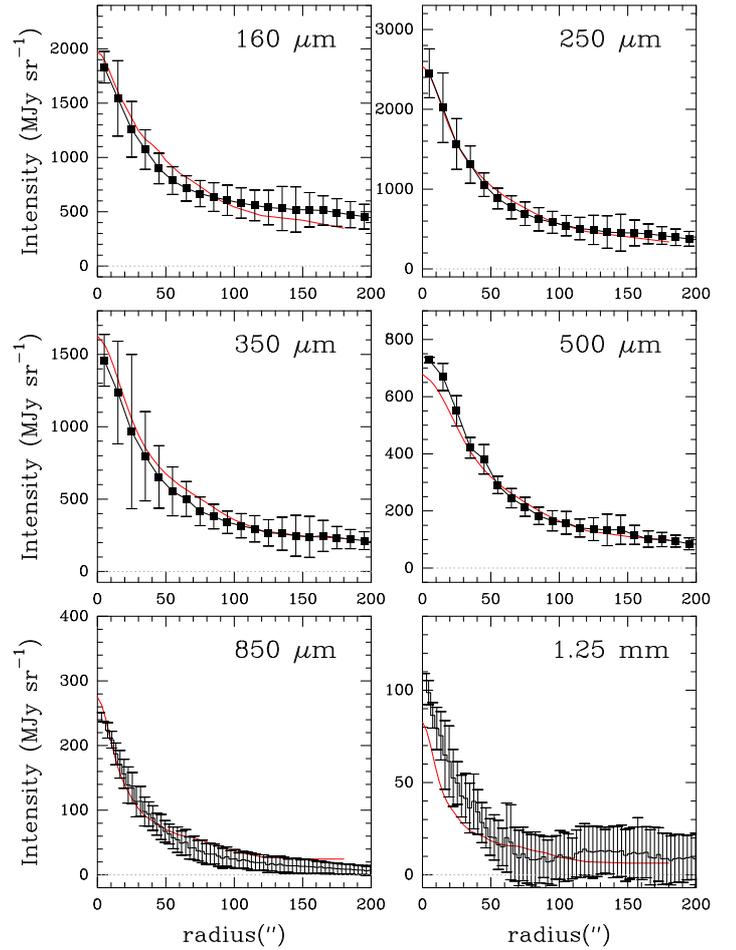}
\caption{Observed continuum surface brightness radial profiles from 160\,$\mu$m to 1.3\,mm (black curve). The model is overplotted in red.} \vspace{-0.1cm}
\label{fig:fit_continuum}
\end{center}
\end{figure}

A first step of the analysis consisted in performing a radial average of the continuum maps. Since the IRAS 16293-2422 protostar 
is located at the North--West from the core, at 
$\alpha$ = 16$^h$32$^m$22.73$^s$, $\delta$ = -24$\degr$28$\arcmin$32.2$\arcsec$ (J2000), 
we removed from the average all the points located 
 less than 80$\arcsec$ from this position. The center of the radial profile is taken at the 
mean of the maxima at 850 $\mu$m and 1.3mm, i.e.  
R.A. = 16$^h$32$^m$28.8$^s$,  Dec. = $-24\degr$29$\arcmin$4$\arcsec$ (J2000), 
which is offset by $\sim$3$\arcsec$ with respect to the position where NH and ND were observed, well within the
beams of the heterodyne observations.
The radial averages of the continuum maps are reported in Fig. \ref{fig:fit_continuum}.

Since the computation of the radiative transfer
is time-consuming, it was not possible to explore the complete parameter space. We therefore tried several types of density profiles before
adopting a piecewise powerlaw profile, as in \citet{Daniel:2013cb}, with a
constant density value from $r=0$ to $r=r_0$ and a power-law index of $\alpha$ beyond $r_0$ and up to $r_{\rm{out}}$. 
We fixed the innermost limiting radius at $r_0$ = 4$\arcsec$, as in \citet{Stark:2004hz}, and the external radius at $r_{\rm{out}}$ = 350$\arcsec$.
Moreover, we assumed a value of $\kappa_{1300}$ = 0.005 cm$^{2}$ g$^{-1}$ for the dust absorption coefficient at 1.3 mm \citep{Ossenkopf:1994tq}. This density structure can well account for the
continuum emission profiles while keeping the number of free parameters at a minimum. 
The only free parameters of the model are the central density $n_0$, the slope $\alpha$, the dust emissivity spectral index $\beta$, and the dust temperature radial profile. In order to further reduce the number of models, we opted to
fix the dust temperature profile and then run a grid of models with free parameters $n_0$, $\alpha$ and $\beta$.  
We started with the most simple assumption of a uniform temperature across the core. 
Doing so, the most acceptable model is obtained for T$_d$ $\sim$ 14\,K.
For this dust temperature, the model parameters are in the range $ 1.3 < \alpha < 1.7 $, 
$1.2 < \beta < 1.8 $ and $5 \, 10^6 < n_0 < 2 \, 10^7$ cm$^{-3}$. However, with a uniform dust temperature, 
we could not find a model that would fit satisfactorily the radial intensities at every wavelengths. 
Moreover, the dust exponent of the best models are found around $\beta \sim 1.5$, which is low 
by comparison to the exponent expected for the dust in low mass star forming regions. At this stage, we found that
the usual correlation between the dust temperature and the $\beta$ exponent is present, ie. adopting a lower dust
temperature lead to an increase in the value derived for $\beta$. However, the quality of the models is degraded 
by adopting a lower temperature while keeping the dust temperature uniform. Hence, we made the choice to set
the dust opacity exponent to $\beta = 1.7$ \citep[consistent with the value of 1.6 derived from Planck data - ][]{2014A&A...571A..11P} and then investigated various dust temperature profiles that would enable
to reproduce the observations. As in \citet{Stark:2004hz}, the dust temperature was assumed to increase outwards and 
we found a good solution by adopting a dust temperature that would vary from 11\,K in the core centre up to 16\,K in the outer envelope.
These values are consistent with the temperature obtained from NH$_3$ observations by
\citet{Wootten:1987iy}.
With such a gradient, the best models are obtained 
around $\alpha$$\sim$1.7,  $n_0$$\sim$ $1.4 \, 10^7$ cm$^{-3}$. 

\begin{figure}
\begin{center}
\includegraphics[angle=0,scale=.3]{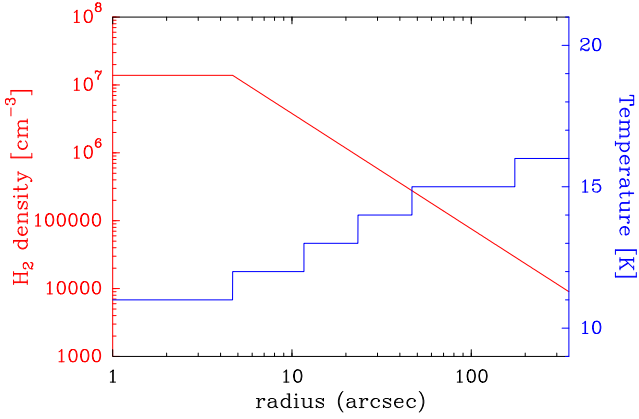}
\caption{H$_2$ density and dust temperature profiles derived from the continuum modeling.}
\label{fig:structure}
\end{center}
\end{figure}

The continuum radial profiles obtained with these parameters are compared to the observations in Fig. \ref{fig:fit_continuum}, 
and the associated physical structure, i.e. H$_2$ density and dust temperature profiles, is reported in Fig. \ref{fig:structure}.
In Fig. \ref{fig:fit_continuum}, the error bars associated with the observations correspond 
to the spread of the observational points introduced by the radial average. As commented in Appendix \ref{appendix:SED}, we expect
additional uncertainties in the absolute flux of the observations due to the calibration of the data, ranging from $\sim15\%$ for the PACS data
to $\sim30\%$ for the ground based one. Hence, these uncertainties can account for the discrepancies between models and observations.

\subsection{Molecules}

The radiative transfer for the molecular line emission was solved using the \texttt{1Dart} code 
\citep{Daniel:2008fy}, which  takes into account the hyperfine structure in the 
excitation calculations. We used the collisional rate 
coefficients of NH and ND with He from \citet{Dumouchel:2012fl}, which were scaled
to account for the difference in reduced mass with H$_2$. Spectroscopic data (frequencies, Einstein spontaneous emission coefficients, level energies, degeneracies) were
originally determined by \citet{Saito:1993dn} and \citet{Klaus:1997ux} for ND and NH, respectively, and were obtained
from the  CDMS\footnote{http://www.astro.uni-koeln.de/cdms/} \citep{Muller:2001ga,Muller:2005ii} database. 
In the case of ND, we found that considering line overlap effects influences the emergent intensities by a factor of $\sim$\,2. 

\begin{figure}
\begin{center}
\includegraphics[angle=0,scale=.3]{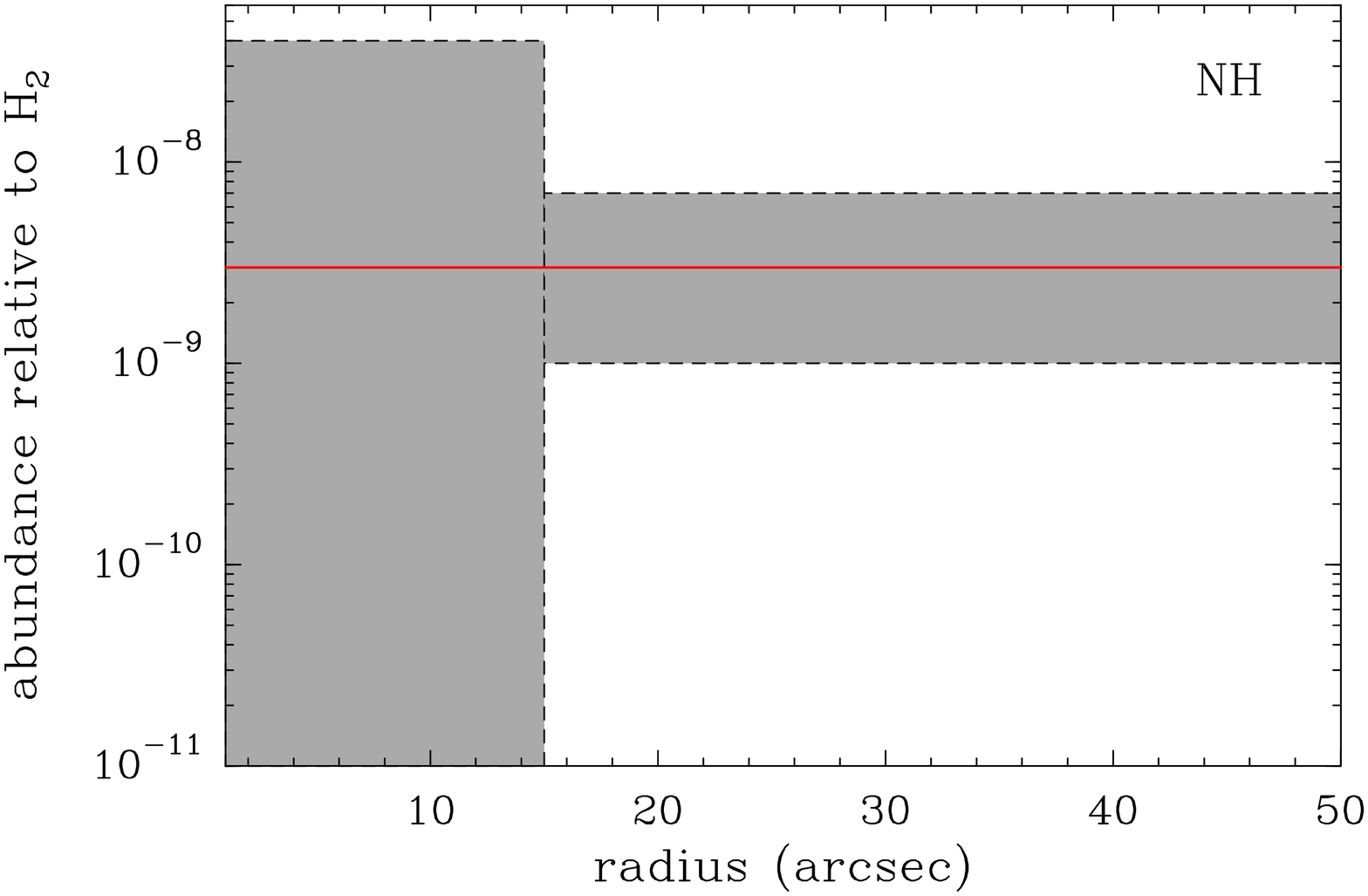}
\includegraphics[angle=0,scale=.3]{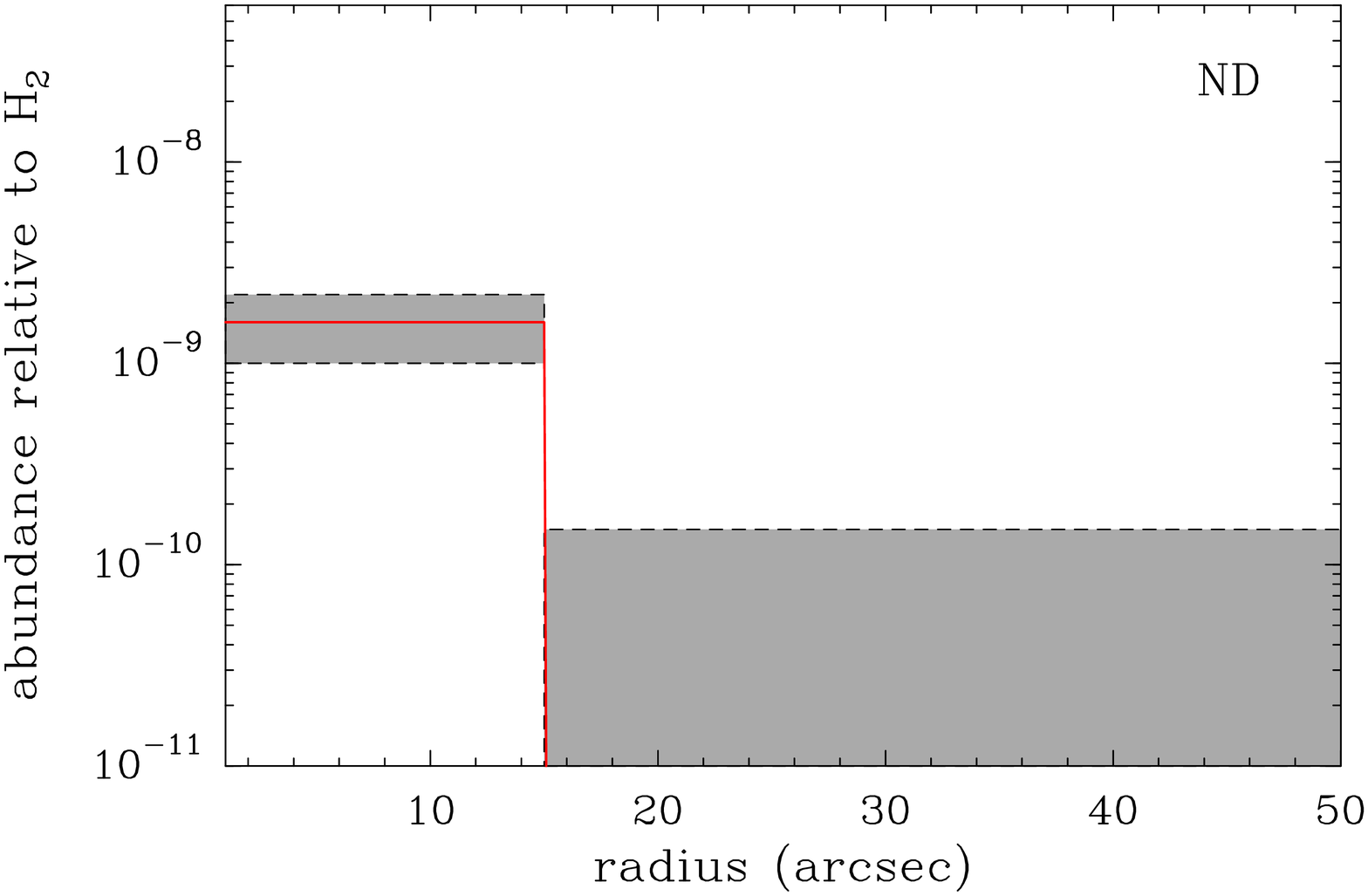}
\caption{NH and ND abundances as a function of radius. The grey area represents the uncertainty on the abundances (from the grey area in Fig.\,\ref{fig:fit_molecules}).} \vspace{-0.1cm}
\label{fig:abundances}
\end{center}
\end{figure}

\begin{figure}
\begin{center}
\includegraphics[angle=0,scale=.4]{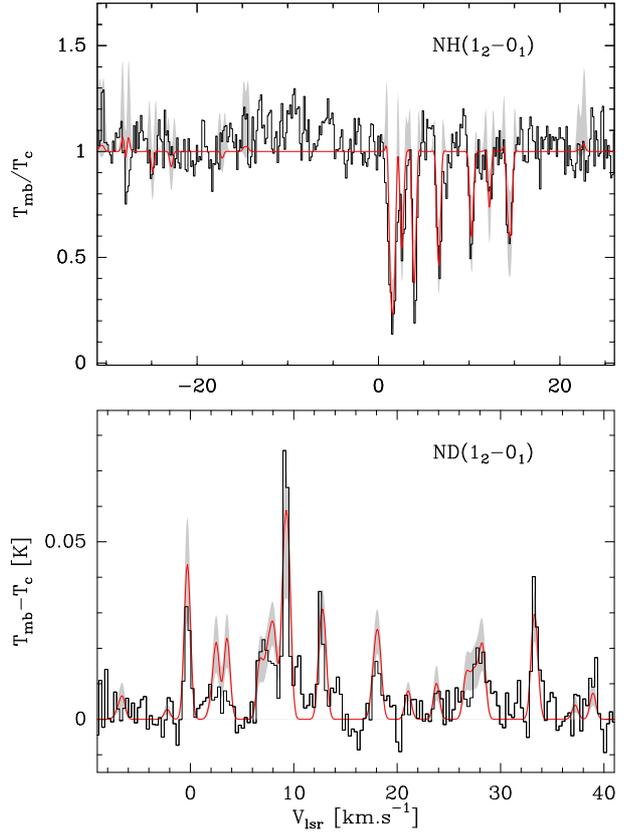}
\caption{Best-fit model (red line) of the non local, non-LTE radiative transfer analysis, superimposed on the observed spectrum (black line). The grey area represents the range of synthetic spectra consistent with the observations. The continuum level $T_c$ has been subtracted from the ND spectrum. The NH spectrum is shown divided by the continuum level $T_c$.} \vspace{-0.1cm}
\label{fig:fit_molecules}
\end{center}
\end{figure}

In order to model the ND spectrum, we first tried to use a constant abundance throughout the envelope.
However, under such an assumption, the resulting fits are rather poor since the abundance needed to produce
enough emission in the lines of lowest opacity leads to self--absorption features in the most opaque lines which 
are by far too pronounced by comparison to the observations. Hence, we introduced two regions in the radial
abundance profile (see Fig. \ref{fig:abundances}). Given that NH and ND were observed toward a single position, 
we do not have enough observational constraints to derive precisely the respective sizes of these two regions.
Indeed, from the modelling, we found that all the models with a central region of radius lower than 60$\arcsec$ would give equivalent fits. Increasing this radius beyond 60$\arcsec$ would lead to self-absorbed profiles in ND, which is not observed. However, the choice of the actual size of the innermost region affects 
the derived value of the ND abundance, and we estimate the uncertainty to be of a factor $\sim$2.
In what follows, we arbitrarily set to $15\arcsec$ the size of the central region. We found a good solution setting the abundance
 to [ND] = 1.6 10$^{-9}$ in this region and derived an upper limit of [ND] = 1.5 10$^{-10}$ outside this radius (Fig.\,\ref{fig:fit_molecules}). 
With the radius of the central region set to $30\arcsec$, the central ND abundance would be [ND] = 1.1 10$^{-9}$ instead. 

For the NH modelling, we first tried a similar abundance profile to ND, scaled by a constant factor.
However, by doing so, we could not reproduce the observations.
In fact, we found that the region of the envelope where the abundance has to be increased in order to fit the observations  
corresponds to the region $r > 15\arcsec$. A reasonable fit to the observations is obtained by setting the NH abundance
to [NH] = $3 \, 10^{-9}$ in this region. Below $15\arcsec$, we arbitrarily kept the abundance of the outer region, i.e. [NH] = 3  10$^{-9}$, 
since the models are not sensitive to this latter value.
Indeed, even in the central region, the densities are not high enough to excite NH, so that the 
NH emission remains within the noise of the observations even for NH abundances of 
 $\sim 4 \, 10^{-8}$.

The comparison between the modelled spectra and the observations is reported 
in Fig. \ref{fig:fit_molecules}, where the red spectrum represents our best fit. Other acceptable fits are 
shown as a grey area. They where determined by varying the ND and NH abundances in the
inner and outer layers while keeping the physical structure (density and temperature profiles) as in the 
best fit (Sect.\,\ref{sec:continuum}). The abundances as well as their uncertainties are shown in Fig.\,\ref{fig:abundances}.

Finally, we stress that the current model most probably represents a simplistic view of the 
actual geometry of the region. Indeed, it was previously observed that the dust emission peak and the deuterium
peak are offset by $\sim$\,15$\arcsec$. We cannot account for such an offset in a 1D model and a 3D modelling 
would thus be required. Obviously, to constrain such a model, it would be necessary to have maps of NH and ND, which
are currently unavailable. Additionally, by examining Fig. \ref{fig:fit_molecules} it is worth noting that the 
comparison between the model and observations shows some discrepancies, especially if one considers the ratio between the 
hyperfine components. In fact, we found that by increasing the gas temperature, we could obtain a better fit to the 
observations. However, if we assume that the dust and gas are thermalized, we would have to decrease the
dust opacity exponent below 1.7 in order to have a model still compatible with the continuum observations discussed in the previous
section. Such a decrease would presumably be unlikely, as the dust emissivity spectral index tends to increase to values above 2 when
the dust temperature decreases below 15\,K \citep[e.g.][]{Paradis:2011jl}. In fact, it is important to keep in mind that the current model 
uses collisional rate coefficients scaled from the He rates. Such a procedure often leads to underestimate the actual rate 
coefficients with H$_2$. As a consequence, if the rate coefficients are indeed underestimated, this could provide an alternate
explanation for the discrepancies between the model and observations, and would not require to modify the gas temperature derived
from the continuum. This subject will be addressed when the collisional rate coefficients with H$_2$ become available.

The column densities of ND and NH were determined by integrating the density 
multiplied by the abundance along the line of sight, and by convolving
with a beam of the appropriate size. 
The values are 
$N(\mathrm{ND})=6.0^{+2.3}_{-2.3}$ 10$^{13}$ cm$^{-2}$
and 
$N(\mathrm{NH})=6.4^{+38.3}_{-3.1}$ 10$^{14}$ cm$^{-2}$.
These values are a factor $\sim$10 higher than the ones derived
from the LTE analysis for NH, and a factor of $\sim$\,6 higher for ND. 

In summary, in order to obtain a reasonable fit of the NH and ND spectra, it is necessary to introduce two regions in the abundance radial profile. The ND emission originates primarily from the inner region and the NH absorption from the outer region. Consequently, the NH abundance in the inner region is poorly constrained and we can only set an upper limit to the ND abundance in the outer region. Adopting a non-LTE approach has allowed us to improve the fit of the ND spectrum, in particular the relative ratios of the
hyperfine components are better reproduced than in the case of the LTE hypothesis.

\section{Discussion} \label{sec:discussion}

The current modelling of the NH and ND spectra observed towards 16293E requires to introduce 
a dichotomy between the two species. From the modelling, the ND emission originates in the innermost part
of the core, while the NH absorption is mostly dominated by a region that surrounds it. Hence, the spectra of both
species are found to be dominated by two regions which are spatially distinct.
The separation between these two regions is further outlined considering the $V_{lsr}$ associated with the 
lines. Indeed, the ND and NH spectra are consistent with $V_{lsr}$ values that differ by $\sim$0.4 km s$^{-1}$.
As a consequence, we conclude that discussing the [ND]/[NH] abundance ratio is not meaningful, or at least largely 
uncertain, in the case of the 16293E prestellar core. At most, we can say that in the region where the ND abundance is constrained, the deuterium  fractionation is larger than 2\%. 
In the outer region, an upper limit for [ND]/[NH] is 20\%. Recently, the formation chemistry of the NH 
isotopologues has been investigated by \citet{HilyBlant:2010jp} and \citet{Roueff:2015cc}. In both studies, the authors conclude
that the main pathway leading to either NH or ND is through dissociative recombination of N$_2$H$^+$ or N$_2$D$^+$ with electrons. The upper and lower limits we infer for [ND]/[NH], in the 16293E core, are consistent with the latest 
theoretical predictions of \citet{Roueff:2015cc}, since their model gives [ND]/[NH]$\sim$10\% at H$_2$ densities $\sim$$10^5$ cm$^{-3}$ and falls below 6\% at  $\sim$$10^4$ cm$^{-3}$.

By considering the current modelling, we note that the parameters we derived in order to describe the core, 
i.e. the H$_2$ density, the radial gas temperature profile or the parameters that describe the dust, 
may be subject to large uncertainties. This is a consequence of strong correlations between 
some model parameters, e.g.
between the dust absorption coefficient and the central H$_2$ density $n_0$, or between the dust temperature
and the dust spectral index $\beta$. The uncertainty, which affects the individual parameters can be outlined by considering
earlier estimates of, e.g., the central core density. As an example, from a model of dust emission maps at 450 
and 850 $\mu$m, \citet{Stark:2004hz} estimated the H$_2$ density to $1.6 \, 10^6$ cm$^{-3}$ in the central 1000 AU, while
our current estimate is a factor of 10 larger. Our estimate of the central density is however consistent with the 
qualitative conclusion reached by \citet{Lis:2002ft}, from observation of a ND$_2$H line of high critical density. Indeed,
\citet{Lis:2002ft} concluded that 16293E and L1544 should have comparable central densities, that would be typically 
a factor of two higher than the density in B1. Hence, though the central density value derived from the current modeling is high,
it is consistent with the latest estimate of the H$_2$
density in L1544 and B1, respectively estimated to $\sim$$10^7$ within a central radius of 500 AU \citep{Keto:2014fo} and $3 \, 10^6$ cm$^{-3}$ within a radius of 1175 AU \citep{Daniel:2013cb}.
Finally, given these uncertainties, we modelled the NH and ND spectra with alternate density and temperature profiles, where
the central density was, at most, typically varied by a factor 3 according to the model discussed earlier.  The temperature was varied
by a few kelvins. The absolute
abundances of NH and ND with respect to H$_2$ change accordingly (i.e. when the density is decreased by a given factor, the abundances
increase by the same factor), but both the abundance profiles as well as the [ND]/[NH] abundance ratios remain unchanged, so that
the dichotomy between the NH and ND emission is still present in these models.

Prior to this work, ND was only detected once, in the direction of the IRAS 16293-2422 
protostar \citep{Bacmann:2010it}. Its lines, along with the main isotopologue lines, were detected in
absorption and it was deduced that both molecules would trace the cold envelope that surrounds the protostar.
The abundance ratio between the two isotopologues was estimated to lie in the range [ND]/[NH]$\sim$30-100\%.
This range of values is larger than the upper limit of 20\% we deduced for the extended component traced by NH, in direction of the 16293E prestellar core, but agrees with the lower limit we derived for the region where ND is detected.
Finally, in the direction of the IRAS 16293-2422 protostar, it was found that some deuterated species, like HDO, would trace an external foreground layer \citep{Coutens:2012hx,Wakelam:2014hi}, used to reproduce Herschel/HIFI CH observations \citep{Bottinelli:2014cw}.
In the current case, we cannot confirm nor discard the presence of this foreground layer in the direction of the 16293E core. 
Our data do not enable us to conclude about the origin of the velocity shift between the ND and NH spectra i.e. whether NH absorption arises in a foreground layer component, whether the denser gas is blueshifted because of its interaction with the outflow as suggested by \citet{Lis:2002ft}, or whether the velocity difference could be due to gravitational contraction.

The fact that hydrogenated and deuterated species come from distinct physical regions, in L1689N,
was previously recognized by \citet{Lis:2002jz}.
However, such a distinction would not necessarily be recognized for the molecules
which are observed with insufficient velocity resolution. In the present case, because of the difference in excitation for the two species
(the critical densities for the NH $1_2-0_1$ transitions are a factor of ten higher than those for the ND $1_2-0_1$ transition), NH does not emit in the central region where ND emission originates and the emission of hydrogenated and deuterated species are dominated by distinct physical regions.  For a better determination of deuterium fractionation, lines with similar critical densities should be used for both the main species and its deuterated counterpart.
 
\begin{acknowledgements}
Nicolas Billot is acknowledged for his help in reducing PACS data in HIPE. We also thank Alexandre Faure for enlightening discussions about collisional excitation of hydrides. Support for this work was provided by NASA (\emph{Herschel} OT funding) through an award issued by JPL/ Caltech.  P.C. acknowledges financial support of the European Research Council (ERC; project PALs 320620). This research has made use of data from the Herschel Gould Belt survey (HGBS) project (http://gouldbelt-herschel.cea.fr). The HGBS is a Herschel Key Programme jointly carried out by SPIRE Specialist Astronomy Group 3 (SAG 3), scientists of several institutes in the PACS Consortium (CEA Saclay, INAF-IFSI Rome and INAF-Arcetri, KU Leuven, MPIA Heidelberg), and scientists of the Herschel Science Center (HSC). This work has been supported by the Agence Nationale de la Recherche (ANR-HYDRIDES), contract ANR-12-BS05-0011-01.

\end{acknowledgements}

\bibliographystyle{aa}
\bibliography{biblitex}

\begin{appendix}
\section{Rest frequencies and relative intensities of observed NH and ND hyperfine components.}\label{sec:frequencies}

Table\,\ref{nhfreq} and \ref{ndfreq} list the rest frequencies of the observed hyperfine components in NH and ND. The relative
intensities have been normalised so that their sum is equal to one. The data
are adapted from the CDMS catalogue \citep{Muller:2001ga,Muller:2005ii}.

\begin{table}
  \caption{NH frequencies and relative intensities of the hyperfine components of the 974.5\,GHz transition.}
\label{nhfreq}
\begin{tabular}{cc}
\hline\hline
Rest frequency & Relative intensity \\
(MHz) & \\
\hline
974315.58 & 3.00 10$^{-5}$ \\
974342.57 & 1.70 10$^{-4}$ \\
974354.64 & 9.00 10$^{-5}$ \\
974410.56 & 1.71 10$^{-3}$ \\
974411.39 & 4.92 10$^{-3}$ \\
974436.35 & 4.38 10$^{-2}$ \\
974437.54 & 3.67 10$^{-2}$ \\
974444.04 & 3.47 10$^{-2}$ \\
974450.44 & 4.61 10$^{-2}$ \\
974462.22 & 9.68 10$^{-2}$ \\
974471.00 & 1.63 10$^{-1}$ \\
974475.41 & 6.48 10$^{-2}$ \\
974478.38 & 2.67 10$^{-1}$ \\
974479.34 & 1.73 10$^{-1}$ \\
974531.32 & 4.98 10$^{-3}$ \\
974539.82 & 6.20 10$^{-3}$ \\
974558.07 & 9.43 10$^{-3}$ \\
974564.78 & 1.47 10$^{-2}$ \\
974574.43 & 2.35 10$^{-2}$ \\
974583.03 & 4.97 10$^{-3}$ \\
974607.78 & 3.49 10$^{-3}$ \\
\hline
  \end{tabular}
\end{table}

\begin{table}
  \caption{ND frequencies and relative intensities of the hyperfine components of the 522.0\,GHz transition.}
\label{ndfreq}
\begin{tabular}{cc}
\hline\hline
Rest frequency & Relative intensity \\
(MHz) & \\
\hline
  521982.65 & 1.00 10$^{-5}$ \\
  521992.18 & 2.20 10$^{-4}$ \\
  522001.05 & 6.50 10$^{-4}$ \\
  522008.98 & 1.57 10$^{-3}$ \\
  522010.65 & 1.93 10$^{-3}$ \\
  522010.65 & 1.60 10$^{-3}$ \\
  522018.95 & 1.76 10$^{-3}$ \\
  522025.72 & 3.19 10$^{-3}$ \\
  522028.73 & 1.95 10$^{-3}$ \\
  522035.57 & 2.23 10$^{-2}$ \\
  522035.57 & 1.63 10$^{-2}$ \\
  522035.57 & 2.96 10$^{-2}$ \\
  522044.32 & 2.73 10$^{-2}$ \\
  522045.27 & 6.01 10$^{-3}$ \\
  522046.00 & 1.44 10$^{-2}$ \\
  522047.06 & 1.76 10$^{-2}$ \\
  522052.07 & 5.04 10$^{-3}$ \\
  522056.83 & 1.61 10$^{-2}$ \\
  522061.94 & 8.01 10$^{-2}$ \\
  522062.72 & 8.95 10$^{-3}$ \\
  522071.29 & 1.24 10$^{-1}$ \\
  522071.29 & 5.57 10$^{-2}$ \\
  522077.37 & 1.32 10$^{-1}$ \\
  522077.37 & 2.00 10$^{-1}$ \\
  522079.53 & 8.93 10$^{-2}$ \\
  522080.30 & 2.31 10$^{-2}$ \\
  522080.64 & 2.26 10$^{-2}$ \\
  522081.82 & 1.93 10$^{-2}$ \\
  522087.35 & 2.22 10$^{-2}$ \\
  522089.14 & 2.05 10$^{-2}$ \\
  522090.05 & 4.52 10$^{-3}$ \\
  522094.03 & 2.22 10$^{-2}$ \\
  522097.43 & 1.63 10$^{-3}$ \\
  522104.99 & 2.66 10$^{-3}$ \\
  522105.87 & 7.90 10$^{-4}$ \\
  522115.01 & 8.10 10$^{-4}$ \\
  522125.61 & 4.00 10$^{-4}$ \\
  522129.42 & 8.70 10$^{-4}$ \\
  522141.23 & 1.20 10$^{-4}$ \\
  522158.81 & 1.00 10$^{-5}$ \\
  \hline
  \end{tabular}
\end{table}

\section{Spectral energy distribution of 16293E} \label{appendix:SED}

\begin{figure}
\begin{center}
\includegraphics[angle=0,scale=.48]{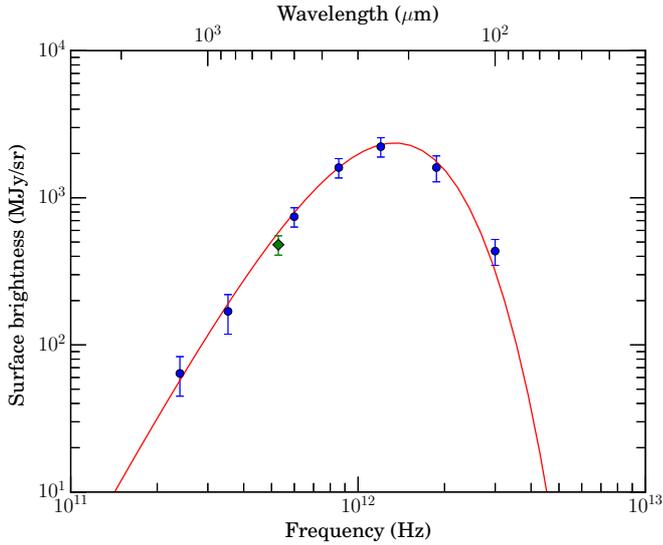}
\caption{\label{sed}Spectral energy distribution of 16293E. The round blue dots are measurements from the continuum maps (PACS at 160$\mu$m, SPIRE at 250, 350, and 500\,$\mu$m, SCUBA-2 at 850 $\mu$m and MPIfR bolometre at 1.3\,mm), the green diamond is the continuum measured on the ND spectrum. The red line is a modified blackbody law with a temperature of 14\,K, an H$_2$ column density of 1.66\,10$^{23}$\,cm$^{-3}$, and a dust emissivity spectral index $\beta$ of 1.7.}
\end{center}
\end{figure}

From the continuum data described in section \ref{sec:obs}, we derived a spectral energy distribution for the position of integration in ND and NH (Fig.\,\ref{sed}). The continuum data were first smoothed to the spatial resolution of the coarsest map (Herschel/SPIRE at 500\,$\mu$m), i.e. 36$\arcsec$. To this effect, the PACS 160\,$\mu$m, and SPIRE 250\,$\mu$m, 350\,$\mu$m and 500\,$\mu$m were convolved with the appropriate kernels as described in \citet{Aniano:2011gt}. The SCUBA-2 map at 850\,$\mu$m and MPIfR bolometre map at 1.3\,mm were convolved using gaussian kernels of FWHM = 33.2$\arcsec$ and FWHM = 34.3$\arcsec$, respectively. The specific intensities given are those measured at the position of the ND integration. Error bars on the absolute intensities were estimated to be 15\% for PACS data, 20\% for SPIRE data and 30\% for ground-based data. With these values, we have tried to take into account the uncertainties due to large-scale diffuse emission components not related to the cloud, unknown thermal emission of the telescope, and  filtering of the extended emission for the maps coming from ground-based telescopes, as well as the calibration errors quoted in Herschel's reports on extended emission calibration \citep{Paladini:2012up,Griffin:2013kq}. The continuum measured on the ND spectrum (which has a similar angular resolution of 41$\arcsec$) is also plotted on the spectral energy distribution, and is consistent with the specific intensities measured from the continuum maps, within the uncertainties. In addition, we plotted a modified blackbody law accounting for the observed continuum emission at the various frequencies, following:
\begin{displaymath}
F_\nu = \kappa_0 \left(\frac{\nu}{\nu_0}\right)^{\beta} \mu m_{\rm H} N_{\rm H_2} B_{\nu}(T_{\rm d})
\end{displaymath}
where $\kappa_0$ is the dust emissivity coefficient at frequency $\nu_0$, $\mu$m$_{\rm H}$ the mean molecular mass,  $N_{\rm H_2}$ the total H$_2$ column density and B$_{\nu}(T_{\rm d})$ the black body thermal dust emission at temperature $T_{\rm d}$. In Fig.\,\ref{sed}, we took $\kappa_0=0.005$\,cm$^{2}$g$^{-1}$ at $\nu_0=250$\,GHz, $\beta=1.7$, $\mu=2.8$ \citep{Kauffmann:2008jj}, $N_{\rm H_2}=1.66\,10^{23}$\,cm$^{-2}$ and T$_{\rm d}=14$\,K. 
At the frequency of the NH observation, we estimate the continuum emission from the fit of the spectral energy distribution to be $\approx 2000$\,MJy/sr, i.e. $\approx 0.07$\,K averaged over a beam of $\sim$36$\arcsec$. For the purpose of comparison, 
for the same beam size, the H$_2$ column density of the model discussed in Sec. 4 is $N_{\rm H_2}=1.5\,10^{23}$\,cm$^{-2}$.

\end{appendix}

\end{document}